\documentclass[conference]{IEEEtran}
\IEEEoverridecommandlockouts

\makeatletter
\def\ps@IEEEtitlepagestyle{%
  \def\@oddfoot{\mycopyrightnotice}%
  \def\@oddhead{\hbox{}\@IEEEheaderstyle\leftmark\hfil\thepage}\relax
  \def\@evenhead{\@IEEEheaderstyle\thepage\hfil\leftmark\hbox{}}\relax
  \def\@evenfoot{}%
}
\def\mycopyrightnotice{%
  \begin{minipage}{\textwidth}
  \centering \scriptsize
  Copyright~\copyright~2024 IEEE. Personal use of this material is permitted. 
  Permission from IEEE must be obtained for all other uses,
  in any current or future media, including\\reprinting/republishing this material
  for advertising or promotional purposes, creating new collective works,
  for resale or redistribution to servers or lists, or reuse of any copyrighted
  component of this work in other works by sending a request to pubs-permissions@ieee.org.
  \end{minipage}
}
\makeatother

\usepackage{cite}
\usepackage{amsmath,amssymb,amsfonts}
\usepackage{graphicx}
\usepackage{textcomp}
\usepackage{xcolor}
\usepackage{multirow}
\usepackage{tabularx}
\usepackage{subfig}
\usepackage{mathtools}
\usepackage{color}
\usepackage{multicol}
\usepackage{listings} 
\usepackage{algorithm}
\usepackage{algpseudocode}
\usepackage{pifont}
\usepackage{hyperref}
\usepackage{orcidlink}

\newcommand{\xmark}{\ding{55}}%

\def\BibTeX{{\rm B\kern-.05em{\sc i\kern-.025em b}\kern-.08em
    T\kern-.1667em\lower.7ex\hbox{E}\kern-.125emX}}
\begin{document}

\title{The Impact of Run-Time Variability on Side-Channel Attacks Targeting FPGAs
\thanks{This work was supported by the European Union’s Chips Joint Undertaking (Chips JU) program under grant agreement No. 101112274 (ISOLDE).}
}

\author{
	\IEEEauthorblockN{Davide Galli\orcidlink{0009-0005-9430-7699},
    Adriano Guarisco,
    William Fornaciari\orcidlink{/0000-0001-8294-730X},
    Matteo Matteucci\orcidlink{0000-0002-8306-6739},
    Davide Zoni\orcidlink{0000-0002-9951-062X}}
	\IEEEauthorblockA{\textit{DEIB},\\
		\textit{Politecnico di Milano}, \\
		Milano, Italy, \\
        \{davide.galli, william.fornaciari, matteo.matteucci, davide.zoni\}@polimi.it,
        adriano.guarisco@mail.polimi.it}
}

\maketitle

\begin{abstract}
To defeat side-channel attacks, many recent countermeasures work by enforcing
random run-time variability to the target computing platform in terms of
clock jitters, frequency, and voltage scaling, and phase shift, also combining the
contributions from different actuators to maximize the side-channel
resistance of the target.
However, the robustness of such solutions seems strongly influenced by several
hyper-parameters for which an in-depth analysis is still missing.

This work proposes a fine-grained dynamic voltage and frequency scaling actuator
to investigate the effectiveness of recent desynchronization countermeasures
with the goal of highlighting the link between the enforced run-time
variability and the vulnerability to side-channel attacks of cryptographic
implementations targeting FPGAs.
The analysis of the results collected from real hardware allowed for a
comprehensive understanding of the protection offered by run-time variability
countermeasures against side-channel attacks. 
\end{abstract}


\section{Introduction}
\label{sec:introduction}
Side-channel attacks~(SCA) represent one of the most
prominent security threats against modern cryptographic implementations.
Starting from the introduction of power~\cite{Kocher99} and
electromagnetic~\cite{Heyszl12} analysis, the design of countermeasures 
has become an increasingly important topic. Notably, side-channel
countermeasures can be organized into two categories: masking and hiding.
Masking countermeasures degrade the side-channel signal-to-noise ratio~(SNR) 
by splitting the sensitive intermediate values into different
shares that are computed independently.
Hiding countermeasures degrade the side-channel SNR 
by increasing the noise component. Unlike masking solutions,
hiding countermeasures do not offer provable security, while the
achievable low resource overhead fueled extensive
research~\cite{Dao21, CGL+24, GCZ24}.
Trace desynchronization represents a relatively cheap and, thus, widely employed
hiding countermeasure that works by invalidating the common attack hypothesis
for which the collected traces should be time-aligned to each other.
Several proposed trace desynchronization techniques leverage the dynamic voltage and
frequency scaling~(DVFS) actuator to randomly change the operating
frequency and voltage of the target computing platform at
run-time~\cite{Yang05, Dao21, Hettwer20}.
Considering deep sub-micron technology nodes, clock jitter represents a process
variability effect that has been observed to contribute severely to trace
desynchronization~\cite{CDP17, Masure20}

Literature demonstrated the effectiveness of
desynchronization countermeasures and highlighted how the attack success rate
degrades with the increase of the desynchronization~\cite{Benadjila20}.
However, there needs to be a clear understanding of
the hyperparameters and their values to maximize the side-channel resistance of
desynchronization countermeasures.

\smallskip\noindent\textbf{Contributions -} Considering modern FPGAs, this work
investigates the link between the trace desynchronization and the side-channel
attack's success rate, leveraging the Pearson correlation coefficient~($\rho$)
quality metric, making two contributions to the state of the art.
\begin{itemize}
\item \textit{Random DVFS module -} 
We introduce a hardware module targeting AMD Xilinx FPGAs that
performs dynamic and fine-grained operating
voltage, frequency, and phase scaling.
Our solution offers fast temporal dynamics in the order of tens of microseconds,
a fine-grained scaling of the operating parameters,
and a continuous operating mode even during voltage and frequency changes.
\item \textit{Real hardware experimental analysis -}
We perform an extensive experimental campaign targeting multiple
AMD Xilinx Artix-7 FPGA instances to assess the effects of trace 
desynchronization through random DVFS, taking into account the phase shift
and inter-chip process variability. The findings are threefold.
First, analog parts in the DVFS actuator can hinder
the attack by introducing low-dynamic components into the power consumption.
However, using a high-pass filter restores most of the side-channel
vulnerabilities.
Second, inter-chip FPGA variability is not sufficient to hide side-channel
leakages.
Third, frequency scaling is the most effective hiding technique of the
evaluated ones.
The robustness of trace desynchronization techniques improves
with the distance between operating frequencies rather than depending on
the number of frequencies used, as suggested in~\cite{Dao21}.
\end{itemize}

\section{Fine-grained random DVFS architecture}
\label{sec:methodology}
This section presents the random dynamic voltage and frequency scaling
architecture~(\emph{rDVFS}) to perform fast and fine-grained actuation.

\begin{figure}[tp]
    \centerline{\includegraphics[width=0.5\textwidth]{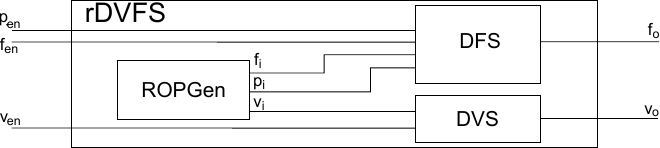}}
    \caption{Architectural view of the proposed random dynamic voltage
    and frequency scaling~(\emph{rDVFS}).}
    \label{fig:rDVFS}
\end{figure}
Figure~\ref{fig:rDVFS} depicts the high-level view of the \emph{rDVFS} architecture.
The \emph{rDVFS} receives three independent control signals, i.e., $f_{en}$,
$p_{en}$, and $v_{en}$, used to selectively enable the random scaling of the
clock frequency~($f_{en}$), the clock phase~($p_{en}$), and the operating
voltage~($v_{en}$), and outputs the generated clock~($f_o$) and the operating
voltage~($v_o$) signals.
Notably, the control signals, i.e., $f_{en}$, $p_{en}$, and $v_{en}$,
allow to selectively force the actuation of a portion or the entire random
operating point. 
The \emph{rDVFS} architecture is made of three sub-modules: \emph{ROPGen},
\emph{DFS}, and \emph{DVS}.

The rest of this section details the architecture of the \emph{ROPGen}
and the \emph{DFS} modules, which represent the on-chip components of
the random DVFS actuator.
Current FPGAs do not offer on-chip DVS actuators, and thus,
we exploit the DVS of the reference board leveraging the SPI/I2C protocol.

\smallskip\noindent\textbf{Random operating point generator (ROPGen) -}
The module implements a free-running true random number
generator~(TRNG)~\cite{GGF+22} to address memories containing the configurations
for each parameter, i.e., clock frequency, clock phase, and operating voltage.
The module's output consists of the random operating point in terms of
frequency, phase, and voltage. The three memories are independently accessed,
and the content of the memories is configured at design time according to the
feasibility requirements described
in the technical specification documents of the considered hardware
components~\cite{datasheet}. The design-time configuration script allows
users to specify boundaries and step increments for frequency and operating
voltage, automatically generating the RTL description for correctly
instantiated memories.
Even if the architecture is generic, we allow up to 1024 frequency
configurations, 8 clock phase configurations, and 128 operating voltage values.
The limited number of phase configurations arises from the practical challenges
of achieving rapid phase shifts for large phase values using the Xilinx
MMCM~\cite{MMCM}. To this end, we obtained large clock phase shifts by
serializing multiple small phase shift actuations.

\begin{figure}[tp]
    \centerline{\includegraphics[width=0.5\textwidth]{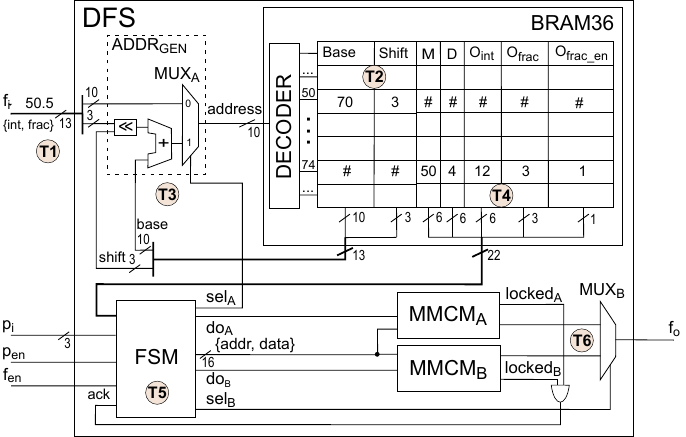}}
    \caption{Architectural view of the proposed dynamic frequency
    scaling~(\emph{DFS}) actuator.}
    \label{fig:dfps}
\end{figure}

\smallskip\noindent\textbf{Dynamic Frequency Scaling (DFS) actuator -}
Figure~\ref{fig:dfps} overviews the architecture of the \emph{DFS} actuator. 
The module takes as inputs the clock frequency~($f_i$) and the clock phase~($p_i$),
as well as two control signals~($f_{en}$ and $p_{en}$) to independently enable
the actuation of the two operating points. The \emph{DFS} outputs the
generated clock signal~($f_o$).
The architecture of \emph{DFS} is organized in three parts:
\emph{i)}~the storage element~($BRAM36$) contains the configuration parameters
for the frequency scaling, 
\emph{ii)}~the pair of Mixed-Mode Clock Modules, i.e., $MMCM_A$ and
$MMCM_B$, represents the actual generators of the clock signal~($f_o$), and
\emph{iii)}~the finite-state machine~($FSM$) and the related address generator
logic~($ADDR_{GEN}$) implement the protocol to perform the frequency and phase
scaling. 

The proposed \emph{DFS} leverages two MMCMs to avoid the clock-gating effect
during the reconfiguration of the clock signal. 
The master MMCM keeps generating the current clock signal while the slave
MMCM undergoes reconfiguration. 
After reconfiguration, their roles switch, ensuring a truly dynamic frequency
scaling actuator.

The FSM module implements the protocol to reconfigure the clock
frequency and phase dynamically.
Inputs consist of clock frequency~($f_i$) and clock phase~($p_i$),
control signals for frequency~($f_{en}$) and phase~($p_{en}$) reconfiguration,
and a signal monitoring the locked status of both MMCMs~($ack$).
Frequency and phase can be reconfigured independently. The $ack$ signal prevents
a new reconfiguration while the previous one is still in progress.

The FSM configures the registers in the memory-mapped MMCM interfaces to change
the clock frequency and phase dynamically. Notably, the phase shift
can be actuated by directly writing the correct MMCM memory-mapped register, while
the frequency reconfiguration involves multiple MMCM registers
for which the corresponding values are obtained from $BRAM36$.
Considering the design of this flexible architecture and the limitations of the
Xilinx MMCMs, the $BRAM36$ can store up to 1024 clock frequency configurations
ranging from 5\,MHz to 800\,MHz due to MMCM feasibility constraints, with a minimum 
step of 0.125\,MHz.

To further illustrate the proposed solution, consider reconfiguring the frequency
of 50.5\,MHz~(see
$T1$-$T6$ in Figure~\ref{fig:dfps}).
Starting from the input frequency value
$f_i=50.5\,MHz$ ($T1$), the integer part initiates the first access to
$BRAM36$~($T2$).
From the obtained record, the \emph{Base} and the \emph{Shift} fields
are used to make the address for the second access to $BRAM36$ containing the
clock frequency configuration for $f_i$. 
In particular, the configuration to obtain an operating
frequency equal to 50.5 is stored at line $74$.
The configuration values $M$, $D$, $O_{int}$,
$O_{frac}$, $O_{frac\_en}$ at line $74$,
are driven to the FSM to configure the slave MMCM registers ~($T5$).
Once the slave MMCM sets its $locked$
signal, the FSM drives $sel_B$ to flip the role of the MMCMs and to propagate
the new frequency through $f_o$~($T6$).

\section{Experimental evaluation}
\label{sec:expEval}
This section discusses the experimental results with the final goal
of highlighting the trace desynchronization parameters that mainly
affect the protection against side-channel attacks (SCA).

\subsection{Experimental setup}
\label{ssec:expEval_setup}
\smallskip
\noindent\textbf{Hardware and software setup -} We implemented the proposed
random DVFS into the NewAE CW305 board~\cite{cw305}. The board features an FPGA
socket that allows changing the mounted FPGA chip.
We employed a 32-bit RISC-V System-on-Chip~\cite{DGC+24, GGF+22} as the reference
computing platform executing an unprotected software implementation of
the AES-128 cryptosystem.
The power traces were collected from the 20~dB-amplified output of the CW305
board using a Picoscope 5244d oscilloscope, sampling at
250~Msamples/s, with a resolution of 12~bits.
The setup features a dedicated hardware signal from the computing platform to
trigger the oscilloscope at the beginning of each encryption.
To prevent measurement pollution due to the ringing effects derived from the
assertion of the trigger signal, we inserted a slight delay between the trigger assertion 
and the actual beginning of the encryption. 

\smallskip
\noindent\textbf{Configurations -}
Our analysis considered the trace desynchronization achieved by scaling
each parameter, i.e., clock frequency, clock phase, operating voltage,
and different FPGA chips, in isolation.
Notably, our \emph{rDVFS} is enabled for the entire encryption.
Thus, a new configuration is requested as soon as one is locked. To this end,
each encryption observes multiple clock frequency and voltage reconfigurations.

For each configuration, Table~\ref{tbl:configurations} reports the boundary
values and the step of the considered parameter.

\begin{table}[t]
	\centering
	\caption{Values in terms of operating voltage, clock frequency, clock phase
    shift, and used chip, for each experimental configuration. 
	}
	\scalebox{0.85}{
		\begin{tabular}{c|c|c|c|c|c|c}
															& \multirow{2}{*}{\textbf{ID}}		& \textbf{Num.} 	 & \textbf{Voltage} 				& \textbf{Frequency}		& \textbf{Phase}		& \textbf{Chip}						\\ 
															&									& \textbf{of steps}  & [V]								& [MHz]						& [°]					& Train\,(Attack)					\\ \hline\hline
			\multirow{4}{*}{\rotatebox{90}{\textbf{Chip}}}	& \multirow{2}{*}{C1}				& \multirow{2}{*}{1} & \multirow{2}{*}{1}				& \multirow{2}{*}{50}		& \multirow{2}{*}{0}	& Artix-7\,100					  	\\			
															&									& 					 & 									& 							&						& (Artix-7\,35)				  		\\ \cline{2-7}
															& \multirow{2}{*}{C2}				& \multirow{2}{*}{1} & \multirow{2}{*}{1}				& \multirow{2}{*}{50}		& \multirow{2}{*}{0}	&  Artix-7\,35					  	\\
															&									& 					 & 									&							&						& (Artix-7\,100)				  	\\ \hline
			\multirow{6}{*}{\rotatebox{90}{\textbf{DVS}}}   & \multirow{2}{*}{V1}     			& \multirow{2}{*}{3} & [0.99; 1.01]						&  \multirow{2}{*}{50} 		& \multirow{2}{*}{0}    & \multirow{18}{*}{Artix-7\,100}   	\\ 
															& 									& 					 & step 0.01						&							&						& \multirow{18}{*}{(Artix-7\,100)} 	\\ \cline{2-6}
                                                			& \multirow{2}{*}{V2}     			& \multirow{2}{*}{2} & \multirow{2}{*}{\{0.75, 1.05\}}  &  \multirow{2}{*}{50} 		& \multirow{2}{*}{0}	& 								  	\\ 
                                                			& 									& 					 &									&							&						& 								  	\\ \cline{2-6}
															& \multirow{2}{*}{V3}     			& \multirow{2}{*}{11}& [0.75; 1.05]		   				&  \multirow{2}{*}{50} 		& \multirow{2}{*}{0}	& 								  	\\ 
															& 									& 					 & step 0.03						&							&						& 								  	\\ \cline{1-6}
			\multirow{2}{*}{\rotatebox{90}{\textbf{DPS}}}	& \multirow{2}{*}{P1}				& \multirow{2}{*}{9} & \multirow{2}{*}{1} 				& \multirow{2}{*}{50} 		& [0; 30]    			& 								  	\\
															&									& 					&									& 							& step  3.75 			& 								  	\\ \cline{1-6}
			\multirow{8}{*}{\rotatebox{90}{\textbf{DFS}}}	& \multirow{2}{*}{F1}				& \multirow{2}{*}{9} & \multirow{2}{*}{1} & [38.375; 39.5] 	& \multirow{2}{*}{0}  	& 									\\ 
															&									& 					 &									& step 0.125				&						& 								  	\\ \cline{2-6}
															& \multirow{2}{*}{F2}    			& \multirow{2}{*}{7} & \multirow{2}{*}{1} 				& [30; 65]			   	   	& \multirow{2}{*}{0} 	& 								  	\\ 
															&									& 					 &									& step  5.0					&						& 								  	\\ \cline{2-6}
															& \multirow{2}{*}{F3}    			& \multirow{2}{*}{6} & \multirow{2}{*}{1} 				& [25; 75]			   	   	& \multirow{2}{*}{0} 	& 								  	\\ 
															&									& 					 &									& step  10.0				&						& 								  	\\ \cline{2-6}
															& \multirow{2}{*}{F3$_{\text{125}}$}&\multirow{2}{*}{401}& \multirow{2}{*}{1} 				& [25; 75]					& \multirow{2}{*}{0} 	& 									\\ 
															&									& 					 &									& step  0.125				&						& 								  	\\ \hline
        \end{tabular}}
	\label{tbl:configurations}
\end{table}

\smallskip
\noindent\textbf{Attack methodology -}
To assess the side-channel vulnerability,
we utilized correlation power analysis (CPA) and template attack (TA),
focusing on the \texttt{SBOX} operation in the initial AES round.
CPA is performed up to 100k traces on the \textit{train} chip of Table~\ref{tbl:configurations},
while the template has been trained considering 1k traces per key value.

Multiple attacks were executed for each scenario and method.
The primary attack used unprocessed traces, while subsequent ones incorporated post-processing,
specifically sample aggregation and a high-pass filter (HPF) with a 125 kHz cutoff
to eliminate low-frequency signal components.

\subsection{Quality metrics}
\label{ssec:expEval_metrics}
Security results are discussed in terms of two quality metrics,
\emph{i)}~Pearson Correlation Coefficient $\rho$
between the power traces and the power estimation is used as an equivalent representation of SNR,
\emph{ii)}~Guessing entropy $GE$ is used to evaluate the effectiveness of the TA,
defined as the average rank position of the correct key among all possible key guesses.

\subsection{Experimental results}
\label{ssec:expEval_results}
Table~\ref{tbl:pvdvfs_results} summarizes
the experimental results.
The \emph{Synch} configuration reports the quality metrics for the reference design with no
desynchronization.
For each scenario~(\emph{ID}), results are reported in terms of $\rho$ values,
number of traces to recover all 16 bytes using \emph{CPA}, and effectiveness of
\emph{TA}.
For each scenario, results are reported with and without the HPF and different levels
of aggregation. Notably, the aggregation can resynchronize the traces by averaging
continuous samples at the risk of a decreased SNR.

\begin{table}[t]
	\centering
	\caption{$\rho$, CPA, and TA comparison between all the configurations, both with and without
	filtering the traces and aggregating up to 1k samples (\checkmark if $\text{GE} = 1$).
	}
	\scalebox{0.95}{
		\begin{tabular}{c|c|c|c c|c c|c c}
							& \multirow{2}{*}{\textbf{ID}}	 & \textbf{Aggregate}              &\multicolumn{2}{c|}{$\rho$}& \multicolumn{2}{c|}{\textbf{CPA}} & \multicolumn{2}{c}{\textbf{TA}}  \\ \cline{4-9}
							&						  		 & \textbf{\textit{n} samples}     & Raw 		& HPF 						& Raw 				& HPF 		 	 & Raw 				& HPF 		    \\ \hline\hline
			\multicolumn{2}{c|}{\multirow{3}{*}{\rotatebox{90}{\textbf{Synch}}}}& 1			   & 0.419      & 0.388 					& 200				& 200			 & \checkmark		& \checkmark    \\ \cline{3-9}
            \multicolumn{2}{c|}{}												& 100		   & 0.587    	& 0.555						& 200				& 200			 & \checkmark		& \checkmark    \\ \cline{3-9}
            \multicolumn{2}{c|}{}												& 1k		   & 0.557   	& 0.360 					& 200				& 500			 & \checkmark		& \checkmark    \\ \hline\hline
			\multirow{2}{*}{\rotatebox{90}{\textbf{Chip}}} & C1     			& 1			   & 0.419      & 0.388						& 200				& 200			 & \checkmark		& \checkmark    \\ \cline{2-9} 
			                                    & C2                        	& 1			   & 0.353		& 0.402   					& 200				& 200			 & \checkmark		& \checkmark    \\ \hline\hline
			\multirow{3}{*}{\rotatebox{90}{\textbf{DVS}}} & V1     				& 1			   & 0.428      & 0.435				    	& 200				& 200			 & \checkmark		& \checkmark    \\ \cline{2-9} 
			                                    & V2                        	& 1			   & 0.089 		& 0.326 					& 10k				& 500			 & \xmark			& \checkmark    \\ \cline{2-9}
			                                    & V3                       		& 1			   & 0.102 		& 0.335						& 10k				& 500			 & \xmark			& \checkmark    \\ \hline\hline									
			\multirow{3}{*}{\rotatebox{90}{\textbf{DPS}}} & \multirow{3}{*}{P1} & 1			   & 0.191  	& 0.185						& 200				& 500			 & \checkmark		& \checkmark	\\ \cline{3-9}
												&                         		& 100		   & 0.220 		& 0.202						& 200				& 500			 & \checkmark		& \checkmark	\\ \cline{3-9}
                                                &                         		& 1k		   & 0.170 		& 0.089						& 500				& 1k			 & \checkmark		& \checkmark	\\ \hline\hline
            \multirow{12}{*}{\rotatebox{90}{\textbf{DFS}}}&\multirow{3}{*}{F1}	&1&0.046& 0.033						& \xmark			& \xmark	 	 & \xmark			& \xmark 		\\ \cline{3-9}
												&                         		& 100		   & 0.215			& 0.184 					& 5k				& 5k			 & \checkmark 		& \checkmark	\\ \cline{3-9}
												&                         		& 1k		   & 0.252			& 0.166 					& 5k				& 5k			 & \checkmark	    & \checkmark	\\ \cline{2-9}
												& \multirow{3}{*}{F2}     		& 1			   & 0.033		& 0.034						& \xmark			& \xmark	 	 & \xmark			& \xmark		\\ \cline{3-9}
                                                 &                         		& 100		   & 0.032		& 0.031						& \xmark			& \xmark	 	 & \xmark			& \xmark		\\ \cline{3-9}
                                                &                         		& 1k		   & 0.031		& 0.018						& \xmark			& \xmark	 	 & \xmark			& \xmark		\\ \cline{2-9}
                                                & \multirow{3}{*}{F3}     		& 1			   & 0.040		& 0.044						& \xmark			& \xmark	 	 & \xmark			& \xmark		\\ \cline{3-9}
                                                &                         		& 100		   & 0.029		& 0.031						& \xmark			& \xmark	 	 & \xmark			& \xmark		\\ \cline{3-9}
                                                &                         		& 1k		   & 0.027		& 0.024						& \xmark			& \xmark	 	 & \xmark			& \xmark		\\ \cline{2-9}
											& \multirow{3}{*}{F3$_{\text{125}}$} & 1	   & 0.035			& 0.033						& \xmark			& \xmark	 	 & \xmark			& \xmark		\\ \cline{3-9}
											&                         		& 100		   & 0.031			& 0.028						& \xmark			& \xmark	 	 & \xmark			& \xmark		\\ \cline{3-9}
											&                         		& 1k		   & 0.028			& 0.024						& \xmark			& \xmark	 	 & \xmark			& \xmark		\\ \hline
        \end{tabular}
		}
	\label{tbl:pvdvfs_results}
\end{table}

\smallskip\noindent\textbf{Inter-chip variability and clock phase -}
The inter-chip process variability on AMD Xilinx Artix-7 FPGAs is too small to prevent
an attack using a model trained on one FPGA instance to target another~(see $C1$
and $C2$ in Table~\ref{tbl:pvdvfs_results}). 
Notably, such results contrast those
reported from analyzing the process variability targeting
ASIC~\cite{Renauld11}, thus motivating the need for separate investigations.
Similarly, dynamic phase shift provokes a negligible degradation in the 
side-channel attacks' success~(see \emph{DPS} in
Table~\ref{tbl:pvdvfs_results}).

\smallskip\noindent\textbf{Operating voltage -}
The attack’s success rate diminishes as the operating voltage range increases.
Considering the \emph{Raw} traces, $\rho$ lowers from
$0.428$ in $V1$ to $0.102$ in $V3$, thus highlighting the benefit of the voltage scaling.
Conversely, using a high-pass filter on the DVS
collected traces shows a non-negligible improvement
of $\rho$, as well the efficacy of \emph{CPA} and \emph{TA}.
The motivation behind such behavior is the low-frequency component
over-imposed by the actuators on the
power consumption. The scaling actuations provoke severe
oscillations in the side-channel signal, thus reducing $\rho$ values and the
effectiveness of the various SCAs.
However, the low dynamic of the over-imposed signal can be removed by applying 
a high-pass filter to the measured traces.

\smallskip\noindent\textbf{Clock frequency -}
Experimental results confirm the effectiveness of dynamic frequency scaling
as desynchronization countermeasures. Similarly to DVS, the side-channel attacks
degrade with the increase in the range of the DFS actuator.
For example, $F1$ and $F3$ employ a frequency range of $[38.375-39.5]$\,MHz and
$[25-75]$\,MHz, respectively. The \emph{CPA} and \emph{TA} results confirm that
$F3$ makes the attacks harder than $F1$. 

Considering the employed post-processing techniques, the high-pass
filter~(\emph{HPF}) fails to improve the SCA. In contrast,
aggregation severally improves the side-channel attacks.
For example, aggregating 100 samples allows \emph{CPA} to
succeed for all key bytes with $5k$ traces in the $F1$ scenario~(see
Table~\ref{tbl:pvdvfs_results}). Notably, none of the considered attacks succeed
in the $F2$ and $F3$ scenarios regardless of the employed post-processing technique,
thus highlighting the effectiveness of severe trace desynchronization
against SCA.

At last, we analyzed the effect of using a large variety of dynamic
frequencies to improve the side-channel resistance as proposed in~\cite{Dao21}.
The frequency scenario $F3_{125}$ considers the step between two consecutive
operating frequencies equal to 125\,kHz.
In addition, $F1$ mimics the effect of clock jitter.
Our experimental results demonstrate that using a high number of
frequencies offers negligible improvements to the side-channel resistance.

\section{Conclusions}
\label{sec:conclusions}
This paper investigated the effectiveness of trace
desynchronization techniques using DVFS targeting AMD Xilinx
Artix-7 FPGAs.
By presenting a random, fine-grained DVFS actuator, we provided a vast
experimental campaign demonstrating that
dynamic phase shift is ineffective as a desynchronization countermeasure.
Moreover, the effect of dynamic voltage scaling is mainly due to the analog
parts of the actuator and can be defeated through a high-pass filter.
Dynamic frequency scaling~(DFS) emerged as the most effective technique to
implement effective desynchronization.
In particular, the resistance against SCA improves as the range of
frequency scaling increases.
Such consideration highlights a critical trade-off between security
and performance, i.e., using DFS for security requires a high variability
in operating frequencies, thus reducing the overall performance and the
execution time predictability.

\bibliographystyle{IEEEtran}
\bibliography{icecs2024}
\end{document}